\documentclass[aps,reprint,onecolumn, showpacs,notitlepage,nofootinbib, superscriptaddress]{revtex4-1}
\usepackage{graphicx}
\usepackage{epstopdf, epsfig}

\usepackage[normalem]{ulem}
\usepackage{cancel}

\AtBeginDocument{\renewcommand{\omega}{V}}

\usepackage{xcolor}
\usepackage{float}

\usepackage[english]{babel}
\usepackage{lipsum}
\usepackage{amssymb,amsfonts,amsmath}

\makeatletter
\newcommand\footnoteref[1]{\protected@xdef\@thefnmark{\ref{#1}}\@footnotemark}
\makeatother

\newcommand{\calH}{\mathcal{H}}
\newcommand{\calU}{\mathcal{U}}
\newcommand{\calX}{\xi}

\renewcommand{\O}{\mathcal{O}}
\newcommand{\uout}{u_o}

\makeatletter
\newcommand{\customlabel}[2]{%
\protected@write \@auxout {}{\string \newlabel {#1}{{#2}{}}}}
\makeatother
\newcommand{\manualsubeq}[1]{\refstepcounter{equation}\customlabel{#1}{\theequation}\customlabel{#1_}{$\alph{equation}$}}

\begin{document}
\title{Theory for the coalescence of viscous lenses}

\author{Walter Tewes}
\affiliation{Physics of Fluids Group, Faculty of Science and Technology, University of Twente, P.O. Box 217, 7500 AE Enschede, The Netherlands
}

\author{Michiel A. Hack}
\affiliation{Physics of Fluids Group, Faculty of Science and Technology, University of Twente, P.O. Box 217, 7500 AE Enschede, The Netherlands
}

\author{Charu Datt}
\affiliation{Physics of Fluids Group, Faculty of Science and Technology, University of Twente, P.O. Box 217, 7500 AE Enschede, The Netherlands
}

\author{Gunnar G. Peng}
\affiliation{Department of Physics and Astronomy, University of Manchester, Oxford Road, Manchester
M13 9PL, UK
}

\author{Jacco H. Snoeijer}  \email{Electronic mail: j.h.snoeijer@utwente.nl}
\affiliation{Physics of Fluids Group, Faculty of Science and Technology, University of Twente, P.O. Box 217, 7500 AE Enschede, The Netherlands
}

\begin{abstract}

	Drop coalescence occurs through the rapid growth of a liquid bridge that connects the two drops. At early times after contact, the bridge dynamics is typically self-similar, with details depending on the geometry and viscosity of the liquid. In this paper we analyse the coalescence of two-dimensional viscous drops that float on a quiescent deep pool; such drops are called liquid lenses. The analysis is based on the thin-sheet equations, which were recently shown to accurately capture experiments of liquid lens coalescence. 
	It is found that the bridge dynamics follows a self-similar solution at leading order, but, depending on the large-scale boundary conditions on the drop, significant corrections may arise to this solution. This dynamics is studied in detail using numerical simulations and through matched asymptotics. We show that the liquid lens coalescence can involve a global translation of the drops, a feature that is confirmed experimentally.  	\end{abstract}

\maketitle

\section{Introduction}\label{Sec.Introduction}
Coalescence of drops is one of the most common capillarity-driven phenomena which can be observed in multiphase fluid dynamics. 
The early-time dynamics of coalescence is dependent on both the viscosity of the drops and their geometry. Different power laws for the growth of the connecting structure (referred to as \textit{neck} or \textit{bridge}) have been found for viscous and inviscid freely suspended drops \citep{Eggers1999,Duchemin2003,Thoroddsen2007,Paulsen2011,Aarts2005}, as well as for sessile drops in the viscous and inviscid limit \citep{Ristenpart2006,HernandezSanchez2012,Eddi2013,Narhe2008,Lee2012}. The study of coalescence phenomena is also relevant for many applications where the underlying substrate of the coalescing drops is a liquid. Some examples are wet-on-wet printing \citep{Hack2018}, emulsions \citep{Shaw2003,Kamp2016},  and lubricant impregnated substrates \citep{Smith2013,Anand2012}.

\begin{figure}
	\centering
	\includegraphics[width=0.85\linewidth]{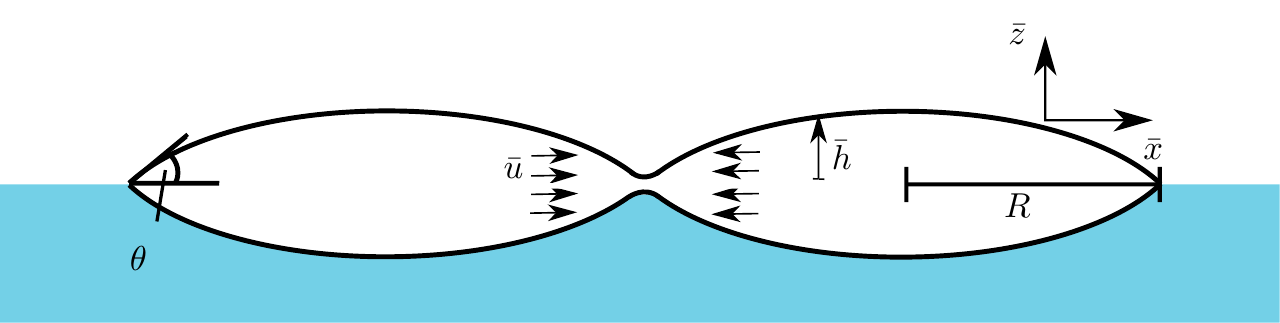}
	\caption{Side-view sketch of two coalescing lenses initial radius $R$ and equilibrium contact angle $\theta$. The drop profile is described by $\bar h(\bar x,\bar t)$, while the minimal height in the neck region $\bar h_0(t)=\bar h(0,\bar t)$. The coalescence velocity $\bar V$ is defined as $\bar{V}=\mathrm{d}\bar{h}_0/\mathrm{d}\bar{t}$.}
	\label{fig:figure1}
\end{figure}

Here, we focus on liquid lenses \citep{deGennes2004}, consisting of liquid drops floating on a quiescent pool of another liquid. This case was studied for Newtonian drops \citep{Burton2007} and liquid crystals \citep{Delabre2010}, where the authors analysed the growth of the bridge in top-view experiments. Recent work considered the coalescence of lenses using side-view experiments \citep{Hack2020}. This perspective is sketched in figure \ref{fig:figure1}, providing a quasi-two-dimensional view of the problem. The experiments revealed a self-similar dynamics of the bridge profiles, with scaling laws for the bridge height $h_0$ with time $t$ that depend on the viscosity of the lenses \citep{Hack2020}. At low viscosity, the dominant balance during coalescence is between surface tension and inertia, and it was found that $h_0\sim t^{2/3}$. At high viscosity, the dominant balance between surface tension and viscosity leads to $h_0 \sim t$.
These scaling laws are the same as those described in the merging of liquid wedges \citep{Billingham2005}. 
Owing to the slender geometry of the drops -- typically the contact angle $\theta$ in figure \ref{fig:figure1} is small -- the coalescence of liquid lenses can be analysed using the thin-sheet equations \citep{Erneux1993,Ting1990}. Using a similarity analysis, the experimentally observed inertial and viscous scaling laws are recovered \citep{Hack2020}. For example, the viscous coalescence speed was found to be $\mathrm{d}\bar{h}_0/\mathrm{d}\bar{t} \approx \bar{V}_0 = 0.5525 \gamma\theta^2/\eta$, 
where $\gamma$ and $\eta$ respectively are the drop surface tension and viscosity (for consistency we here use overbars to indicate dimensional variables). This prediction was found to be in very good agreement with experiments. 

\begin{figure}
	\centering
	\includegraphics[width=0.85\linewidth]{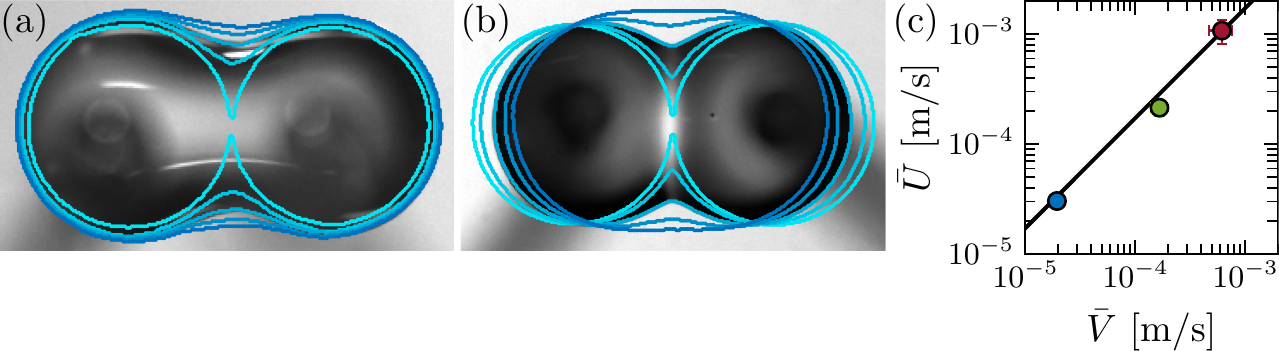}
	\caption{Top-view coalescence of drops in the viscous regime. (a) sessile drops in contact with a substrate, (b) liquid lenses floating on a water bath. During lens coalescence we see a clear inward motion of the drops that does not occur for sessile drops. (c) Comparison of the center of mass velocity $\bar{U}$ and the ``bridge'' coalescence velocity $\bar{V}=\mathrm{d}\bar{h}_0/\mathrm{d}\bar{t}$ for liquid lenses, taken for three oils of different viscosity and nearly identical contact angles (blue: $\eta=$115~Pa$\cdot$s and $\theta=27^\circ$, green: $\eta=$33~Pa$\cdot$s and $\theta=32^\circ$, red: $\eta=$9~Pa$\cdot$s and $\theta=31^\circ$, see \citet{Hack2020} for experimental details). The center of mass velocity $\bar{U}$ is not a small effect, as it is comparable in magnitude in comparison to the coalescence velocity $\bar{V}$.}
	\label{fig:exp}
\end{figure}

The viscous similarity analysis, however, contains a salient feature that remains to be explained: the obtained self-similar velocity profile does not decay at large distance from the thin bridge region, but reaches a finite value. This is rather unusual for problems involving coalescence (or drop breakup, cf. ~\citet{Eggers2015}). Namely, the similarity analysis is typically based on the assumption that the flow remains confined to the scale of the bridge -- at large scale, i.e. the scale of the drop, the flow is usually assumed to vanish. Such is the case for the coalescence of sessile drops as illustrated in figure~\ref{fig:exp}(a). It shows an experimental top-view sequence of merging drops that are in contact with a solid substrate. During the initial growth of the bridge the global features of the drops appear nearly stationary \textemdash{} away from the bridge region one observes only a minute spreading of the drops. Figure~\ref{fig:exp}(b) shows the equivalent top-view sequence for liquid lenses, for which the situation is manifestly different. Clearly, these floating drops do not remain stationary, but their centers of mass exhibit an inward motion as soon as the drops establish contact. This inward motion is not a small effect. Figure~\ref{fig:exp}(c) compares the center of mass velocity $\bar{U}$ (measured in top-view) to the bridge coalescence velocity $\bar{V} = \mathrm{d}\bar{h}_0/\mathrm{d}\bar{t}$ (measured in side-view), showing that the two velocities are proportional. These velocities were obtained using the experimental method described in \citet{Hack2020}. We remark that this inward motion does not at all arise for liquid lenses of very low viscosity -- this is in line with the inviscid similarity solutions, whose velocity rapidly decays away from the bridge~\citep{Hack2020}.

In this paper, we provide a detailed analysis of the coalescence of highly viscous lenses and elucidate the coupling between the inner ``bridge" solution and the global dynamics of the drops. 
{We treat the problem using the two-dimensional thin-sheet equations, reflecting an analysis along  the cross-section shown in figure \ref{fig:figure1}. Such a two-dimensional approximation turned out successful for the geometry of spherical caps \citep{Ristenpart2006,HernandezSanchez2012,Hack2020}, since close to the bridge the length scale in the third dimension, $\O((R \bar{h}_0/\theta)^{1/2})$, is much larger than the horizontal and vertical length scales of the bridge, $\O(\bar{h}_0/\theta)$ and $\O(\bar{h}_0)$.

We demonstrate that, in general, the coalescence velocity exhibits significant logarithmic corrections, $\O(1/ \ln{t})$, where 
 $t\ll 1$ is the (dimensionless) time after coalescence. On the other hand, the thin sheet equations admit an outer solution where the drop's center of mass can migrate freely, closely resembling the motion observed in figure~\ref{fig:exp}(b). In this latter case, 
the corrections to the leading-order result are much smaller. The analysis is confirmed in detail by comparison to time-dependent numerical simulations of the thin-sheet equations.

The article is organized as follows: The governing equations for coalescing lenses are presented in \S\ref{Sec.GoverningEquations}. In \S\ref{Sec.Inner}, we study the inner scale dynamics, near the point of coalescence, followed in \S\ref{Sec.Outer} by an analysis of the outer region, which is where the two different boundary conditions manifest themselves. 
We end with our conclusion and outlook in \S\ref{Sec.Conclusion}.

\section{The viscous thin-sheet equations}\label{Sec.GoverningEquations}

\subsection{Formulation}

Following the approach of \citet{Hack2020}, the process of coalescence is modeled by the two-dimensional viscous thin-sheet equations \citep{Erneux1993,Ting1990}. The underlying approximations are the following: (i) Similar to sessile drops \citep{Ristenpart2006,HernandezSanchez2012} the flow 
in the bridge region 
is quasi-two-dimensional in the early stage of coalescence. (ii) The equilibrium contact angle $\theta$ is small, such that a slender body approximation can be employed. (iii) The influence of the bath on the dynamics is negligible, i.e., free slip boundary conditions can be employed at both interfaces of the 2D lenses. (iv) Due to negligible differences in surface tension between the bath and liquid lens and the liquid lens and air, the liquid lenses are assumed to be symmetrical with respect to the bath-air interface -- though asymmetric surface tensions can actually be mapped to an ``effective" symmetric surface tension (cf. \citet{Hack2020}, Supplementary Material). We note that these assumptions are in accordance to the previous experiments of where the ratio of viscosities of the lenses and the bath is more than a thousand and the corresponding surface tension asymmetry is only about ten percent. The former justifies (iii) and the latter (iv).

The resulting model equations for negligible inertia of the flow, in dimensionless variables, take the form:
\begin{subequations}\label{Eq.GoverningScaled}
\manualsubeq{Eq.MassConservationScaled}
\manualsubeq{Eq.MomentumConservationScaled}
\begin{equation}\tag{\ref{Eq.GoverningScaled}\ref{Eq.MassConservationScaled_},\ref{Eq.MomentumConservationScaled_}}
\dot h + (hu)' = 0,
\qquad
h h''' + 4 (h u')' = 0.
\end{equation}
\end{subequations}
Here, $h(x,t)$ and $u(x,t)$, respectively, are the dimensionless interface height and horizontal velocity, and dots and primes indicate derivatives with respect to $t$ and $x$, respectively. Dimensional variables $(\bar{x},\bar{h},\bar{u},\bar{t})$ are scaled as

\begin{align}\label{eq:scales}
\bar{x}=R x, \quad \bar{h}=\theta R h, \quad \bar{u}=\frac{\gamma}{\eta}\theta u, \quad \bar{t}= \frac{R \eta}{\theta\gamma}t,
\end{align}
where $R$ is the initial lens radius and $\theta$ is the contact angle (cf. figure \ref{fig:figure1}). 
The surface tension is denoted as $\gamma$ and $\eta$ is the viscosity of the liquid inside the lenses. The thin-sheet equations \eqref{Eq.GoverningScaled} correspond to mass and (horizontal) momentum conservation. The latter gives the balance between capillary forces (first term) and viscous forces (second term), while inertia has here been neglected. 

The momentum equation \eqref{Eq.MomentumConservationScaled} can be readily integrated to give the horizontal force balance

\begin{equation}
h h'' - \dfrac{1}{2} h^{\prime 2} + 4 u'h = -\dfrac{1}{2} + F( t).
\label{int_momen_eq}
\end{equation}
The terms on the left-hand side represent the horizontal force transmitted through the thin drop by pressure, surface tension, and viscous stresses, respectively. We have introduced the constant $-1/2$ on the right-hand side corresponding to the total force in the static solution, see \eqref{Eq.Initial} below. Therefore $F(t)$
measures any additional horizontal force that arises during the coalescence dynamics.

\subsection{Two-dimensional numerical simulations}

In order to illustrate the interplay between the dynamics of the small bridge region and the large bulk of the drops, we simulate the coalescence of two-dimensional lenses numerically. Due to symmetry about the point of coalescence $x=0$, we can impose the boundary conditions $h' = u  = 0$ at $x = 0$ and focus on the domain $x \geq 0$. We consider two different sets of outer boundary conditions.

The first case corresponds to the experimentally realised setup of two freely floating lenses. In this case, the length $L(t)$ of each lens decreases with time (as can be seen in figure \ref{fig:exp}) from its initial value $L(t=0) = 2$, and at the edge of the lens we impose the thickness $h(L)$, the contact angle, and conservation of mass:
\begin{equation}\label{Eq.Free}
h(L) = 0,\quad  h'(L) = -1, \quad u(L) = \dot{L}.
\end{equation}
Substitution into the momentum equation \eqref{int_momen_eq} then yields $F(t) \equiv 0$, i.e.\ the horizontal force transmitted through the lens is equal to its initial value at all times. With $F=0$ in \eqref{int_momen_eq}, the condition $h'(L) = -1$ is redundant as it follows from $h(L) = 0$, so we are left with three boundary conditions $h'(0) = u(0) = h(L) = 0$ for the third-order governing equations and a fourth condition $\dot L = u(L)$ that determines the evolution of $L(t)$.

The second case we consider is one where the lenses do not move, due to a symmetry condition being imposed about the centers of the lenses, such as in a periodic array of simultaneously coalescing lenses,
\begin{equation}\label{Eq.Periodic}
h'(1) = u(1) = 0.
\end{equation}
This yields three boundary conditions $h'(0) = u(0) = h'(1) = 0$ on the governing equations and a fourth condition $u(1) = 0$ that determines the unknown $F(t)$.

The key finding is that the different \emph{outer} boundary conditions will lead to different spatial and temporal dynamics also in the \emph{inner} region, at the scale of the bridge. As we shall see, the leading-order solution in the two cases remains the same \textemdash{} however, the coalescence velocity can exhibit significant corrections at 
the next order.

The theoretical initial condition is taken to be the static solution of \eqref{Eq.GoverningScaled} with non-dimensional radius $1$ and contact angle $1$ (corresponding to the dimensional values $R$ and $\theta$, respectively),
\begin{equation}\label{Eq.Initial}
  h(x,t=0) = h_s(x) = \frac{1}{2} x (2 - x),
\end{equation}
but in the numerical simulations a small-scale perturbation $(h_{0i}-x/2)\exp(-x/2h_{0i})$ is added to initialize the coalescence. The perturbation profile is chosen to satisfy the symmetry boundary condition $h'(0) = 0$ and have an initial bridge height $h_{0i} = h_0(t=0)$, which is taken to be $10^{-10}$ unless stated otherwise.

We solve the governing equations \eqref{Eq.MassConservationScaled} and \eqref{int_momen_eq} numerically in Matlab using a finite difference method with a Crank--Nicolson time-stepping scheme. Both the spatial and temporal grids are non-uniform with a relative resolution of $1\%$, and the discretisation is second-order accurate in both space and time. For the periodic lenses \eqref{Eq.Periodic}, the computational domain is $0 \leq x \leq 1$. For the free lenses \eqref{Eq.Free}, where the domain $0 \leq x \leq L(t)$ changes with time, we use a rescaled position variable $\hat{x} = 2x/L(t)$ and solve the rescaled equations on the fixed domain $0 \leq \hat{x} \leq 2$ instead. As explained above, four boundary conditions are imposed on the third-order system of equations, in order to determine the additional unknown $\dot L(t)$ or $F(t)$.

Some snapshots of height and velocity profiles from the simulations are shown in figure \ref{fig:figure3}. The free-floating lenses immediately begin to move towards each other at a constant velocity, and eventually merge into one larger lens with double the volume (and hence $\sqrt{2}$ times the radius and height). The periodic lenses instead flatten out towards a uniform height of $1/3$, determined by the initial volume in the lens. Importantly, and this is one of the central points of the paper, the velocity profiles are significantly different between the two cases, even at very early times, due to the different outer boundary conditions. We will show how this affects the coalescence velocity, $\dot{h}_0$.

\begin{figure}
	\centering
	\includegraphics{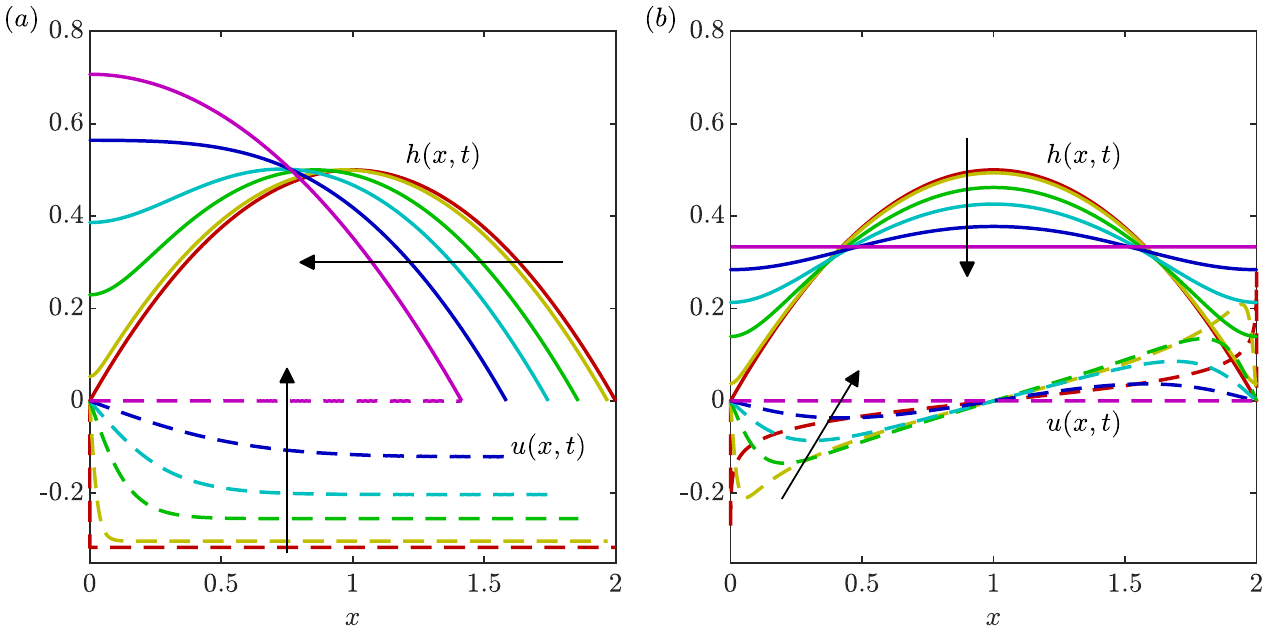}	
	\caption{Numerical height and velocity profiles for the coalescence of ($a$) two free-floating lenses \eqref{Eq.Free} and ($b$) an array of periodic lenses \eqref{Eq.Periodic}, evaluated at the times $t = 10^{-4},\ 0.1,\ 0.5,\ 1,\ 2,\ 10$. The arrows indicate increasing time.
	}
	\label{fig:figure3}
\end{figure}	

\section{The inner region} \label{Sec.Inner}

In order to study the dynamics of the region close to the point of coalescence, we rescale the variables as 

\begin{equation}\label{Eq.InnerDef}
\begin{aligned}
\xi = \dfrac{x}{h_0 ( t )}, \quad
h(x, t) = h_0 ( t ) \, \mathcal{H}( \xi , \tau ), \quad
u(x, t) = \dot{h}_0 ( t ) \, \mathcal{U}(\xi, \tau ),
\end{aligned}
\end{equation}
with $\tau = t$. In choosing the scaling used above, we are motivated by the fact that in the inner region the length scale of importance is the bridge height $h_0$. The resulting governing equations 
become
\begin{subequations}\label{inner_similarity}
\begin{align}
\frac{h_0}{\omega}\dot \calH + \calH - \calX \calH' + (\calH\calU)' = 0, \label{inner_similarity1} \\
 \qquad \calH\calH'' - \frac12 \calH'^2 + 4 \omega \calH \calU' + \frac{1}{2} - F(\tau) = 0,
\label{inner_similarity2}
\end{align}
\end{subequations}
where $\omega = \dot h_0$ is the unknown coalescence velocity that we wish to determine. 
 At $\xi = 0$, the definition $ h(0,t) = h_0(t)$ and the symmetry about $\xi = 0$ yield
 \begin{gather}\label{Eq.InnerBC}
	\calH (0) = 1, \qquad \calH'(0) = \calU(0) = 0. 
\end{gather}

The system of equations \eqref{inner_similarity} has three spatial derivatives and accordingly three boundary conditions \eqref{Eq.InnerBC}. However, the equations contain two unknowns $\omega$ and $F$ that are determined on matching to the outer solution. The matching necessitates that we evaluate $\calH$ and $\calU$ as $\xi \rightarrow \infty$, and from \eqref{inner_similarity} we obtain, on neglecting the time-derivative term,
\begin{gather}
  \calU = \frac{F}{4\alpha \omega}\ln \calX - \beta + \O(\ln\calX/\calX), \quad \calH = \alpha \calX + \O(\ln \calX), \qquad \text{as } \xi\to\infty. \label{Eq.LogFarfield}
\end{gather}
This asymptotics contain two degrees of freedom $\alpha$ and $\beta$ that are determined during the solution process.\footnote{Given that the system is third order in space, there should be a further degree of freedom as $\xi \to \infty$. However, it can be shown by perturbation analysis of \eqref{inner_similarity} (cf. \citet{Eggers2015}), that this degree of freedom appears as a prefactor of a term decaying through an exponential in $\xi$ as long as $\alpha >0$, which is always the case in this coalescence problem. 
The degree of freedom does thus not appear in our leading order asymptotics.} 
This matching will be performed explicitly in \S\ref{Sec.Outer}, but below we already anticipate some of the results necessary to evaluate the inner solutions.

\subsection{The leading-order similarity solution}

The leading-order solution to the foregoing equations is a steady self-similar solution, which was previously obtained by \citet{Hack2020}. The leading-order quantities $\calH_0(\xi)$, $\calU_0(\xi)$ and $\omega_0$  are independent of $\tau$, so that  \eqref{inner_similarity} reduces to

\begin{subequations}
\begin{align}
\calH_0 - \calX \calH_0' + (\calH_0\calU_0)' = 0, \label{inner_similarity1_lo} \\
 \qquad \calH_0\calH_0'' - \frac12 \calH_0^{'2} + 4 \omega_0 \calH_0 \calU_0^{'} + \frac{1}{2}  = 0,
\label{inner_similarity2_lo}
\end{align}
\end{subequations}
where we have anticipated that $F(\tau) \ll 1$ \textemdash this indeed is true as we see later. The boundary conditions required to evaluate these quantities are given by \eqref{Eq.InnerBC}, complemented by $\calH'_0 = 1$ as $\xi \rightarrow \infty$ ($\alpha = 1$) set by the leading-order outer solution given in \eqref{Eq.Initial}. These equations (and also the higher-order equations) are solved numerically in Mathematica using a shooting method on the domain $0 \leq \calX \leq 10^5$. We find the leading-order coalescence and far-field translational velocities to be \citep{Hack2020} 
\begin{equation}\label{Eq.LeadingResult}
 \omega_0 = 0.5525, \qquad \calU_0 \to -\beta_0 = -0.5734 \text{ as } \xi\to\infty.
\end{equation}

\begin{figure}
	\centering
	\includegraphics{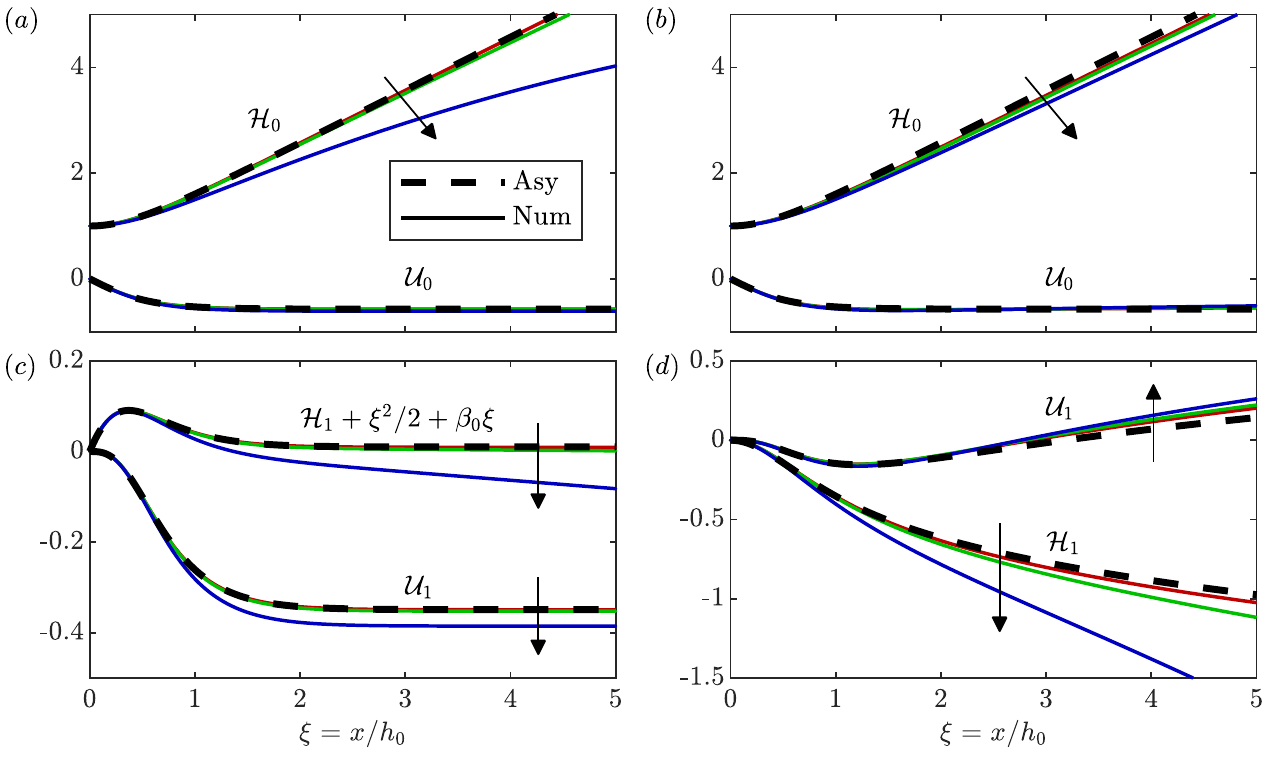}
	\caption{Similarity solution profiles --- leading order \eqref{Eq.LeadingResult} and first correction (\ref{Eq.CorrectionResultFree},\ref{Eq.CorrectionResultPeriodic}) --- for ($a,c$) free-floating lenses and ($b,d$) periodic lenses. Rescaled numerical results are shown for comparison, evaluated at ($a,c$) $t = 10^{-3},\ 10^{-2},\ 10^{-1}$ and ($b,d$) $t = 10^{-4},\ 10^{-3},\ 10^{-2}$. The numerical height profile is transformed as $\calH_{0,{num}} = h/h_0$ and $\calH_{1, num} = (h/h_0 - \mathcal{H}_{0, asy})/\epsilon$ where $\epsilon = h_0$ for free lenses and $\epsilon = F$ for periodic lenses, and the velocity is transformed similarly. (In order to resolve the $\O(10^{-3})$ corrections in ($c$), the numerical simulation was performed with a resolution of $0.1\%$, which restricted $h_{0i}$ to the larger value $10^{-6}$.) The arrows indicate increasing time.
}
	\label{fig:figure4}
\end{figure}

\subsection{Next-order corrections}

The solutions $\calH_0$ and $\calU_0$ are plotted in figure \ref{fig:figure4}($a$,$b$), where they are compared to the numerical solutions. The two columns correspond to the two types of boundary conditions, free-floating lenses and periodic lenses. For both cases and excellent agreement is found at early times after coalescence. However, the velocity profiles in figure \ref{fig:figure3}(a,b) revealed large differences between these two situations. While these differences do not turn up in the leading order inner solutions  $\calH_0$ and $\calU_0$, they impact the next order corrections. As we will see, this also has a strong effect on the evolution of the coalescence velocity $V$. We therefore now address the inner solution beyond the leading order.

\subsubsection{The case $F(\tau) \equiv 0$}\label{Sec.InnerFree}

In the case of free-floating lenses, for which $F(\tau)\equiv0$, it turns out that we can set up a consistent expansion based on the expansion parameter $h_0(\tau)$. On substituting the expansions
\begin{gather}
  \{\calH(\tau),\calU(\tau),\omega(\tau), \beta(\tau)\} = \{\calH_0,\calU_0,\omega_0, \beta_0\} + h_0(\tau)\,\{\calH_1,\calU_1,\omega_1, \beta_1\} + O(h_0^2)
\end{gather} 
(where the dependence of $\calH$ and $\calU$ on $\xi$ is understood) into equations \eqref{inner_similarity}--\eqref{Eq.InnerBC} we obtain  at $\O(h_0)$, 
\begin{subequations}\label{Eq.SimFreeEq}
\begin{gather}
  2\calH_1 - \calX\calH_1' + (\calH_0\calU_1 +  \calH_1\calU_0)' = 0, \label{Eq.SimFreeEq1} \\
	\calH_0\calH_1'' + \calH_1 \calH_0'' - \calH_0'\calH_1' + 4(\omega_0\calH_0 \calU_1' + \omega_0\calH_1\calU_0' + \omega_1\calH_0\calU_0') = 0, \label{Eq.SimFreeEq2} \\
	\calH_1(0) = \calH_1'(0) = \calU_1(0) = 0. \label{Eq.SimFreeEq3}
\end{gather}
\end{subequations}

The differential equations for $\calH_1$ and $\calU_1$ are of third order with three boundary conditions, so one further condition is required in order to determine the unknown $\omega_1$. This is obtained from a matching with the outer region. As $\calX \to \infty$, the generic solutions to \eqref{Eq.SimFreeEq} are quadratic, $\calH_1 \propto \calX^2$, and so will need to match the second derivative $h_s''(0) = -1$ (an $\O (h_0)$ quantity when expressed in inner variables) of the outer solution. We thus impose $\calH_1'' \to -1$ as $\calX \to \infty$, and obtain the numerical solutions plotted in figure \ref{fig:figure4}($c$), with the coefficients
\begin{equation}\label{Eq.CorrectionResultFree}
  \omega_1 = -0.7625, \quad \calU_1 \to -\beta_1 = -0.3492 \text{ as } \calX \to\infty.
\end{equation} 

\subsubsection{The case of non-zero $F(\tau)$}\label{Sec.InnerPeriodic}

When the horizontal force $F(\tau)$ is non-zero, the next-order corrections are at $\O(F)$. As we shall see in the next section, these corrections dominate the  $\O( h_0)$ corrections that arise due to time evolution of the outer solution, since $F$, though unknown yet, is evaluated to be $\O(1/\ln(h_0))$. 
For this reason, we now expand the solution in terms of $F$, i.e.
\begin{equation}
\left\{ \calH(\tau), \calU(\tau), \omega(\tau), \beta(\tau) \right\} = \left\{ \calH_0, \calU_0, \omega_0, \beta_0  \right\} + F(\tau)  \left\{ \calH_1, \calU_1, \omega_1, \beta_1  \right\} + \O(F^2).
\end{equation}
After substituting into the governing equations, we obtain, at $\O(F)$,
\begin{subequations}
\begin{gather}
\calH_1 - \calX \calH_1' + (\calH_0 \calU_1 + \calH_1 \calU_0)' = 0, \\
\calH_0 \calH_1'' + \calH_1 \calH_0'' - \calH_0' \calH_1' + 4 (\omega_0 \calH_0 \calU_1' + \omega_0 \calH_1 \calU_0' + \omega_1 \calH_0 \calU_0') = 1, \\
\calH_1'(0) = \calU_1(0) = \calH_1(0) = 0.
\end{gather}
\end{subequations}
Since the corrections to the outer solution at $\O(h_0)$ are to be neglected when matching the inner and outer solutions at $\O(F)$, we now impose the condition $\calH_1' \to 0$ as $\calX \to \infty$.
On solving these equations numerically, we obtain the solution profiles plotted in figure \ref{fig:figure4}($d$) and the coefficients
\begin{equation}\label{Eq.CorrectionResultPeriodic}
  \omega_1 = -0.6227, \quad \beta_1 = 0.7079.
\end{equation}

As $F = \O(1/\ln h_0)$ will typically not be very small, it is useful to calculate the second-order $\O(F^2)$ correction, which includes contributions proportional to $F^2$ as well as contributions proportional to $(h_0/\dot h_0)\dot F$. We perform this calculation in appendix \ref{Sec.Appendix}.

\section{The outer region and matching}\label{Sec.Outer}

We now turn to the outer solution, and calculate the corrections to the initial static profile $h_s$ \eqref{Eq.Initial} that are generated by the coalescence in the bridge region.
Time integration of the evolution equation \eqref{Eq.MassConservationScaled} from this initial condition yields 
\begin{equation}\label{Eq.EvolutionIntegral}
h(x,t) = h_s(x) - \int_0^t \left(u(x,\hat t) h(x,\hat t) \right)'\,\mathrm{d}\hat t.
\end{equation}
The outer solution, where $h,u,x = \O(1)$, is thus given by the initial profile $h_s(x)$ with an $\O(t) = \O(h_0)$ correction. \\Integration of the momentum equation \eqref{int_momen_eq} using $h = h_s + \O(h_0)$ then yields the outer-region velocity profile
\begin{equation}\label{Eq.OuterVelocity}
\uout = \dfrac{F(t)}{4} \ln{\dfrac{x}{2-x}} + B(t) + \O(h_0).
\end{equation}
The coefficients $B(t)$ and $F(t)$ are determined by boundary conditions and by matching to the inner solution.

\subsection{The case of freely floating lenses}

For coalescing free lenses, the boundary conditions \eqref{Eq.Free} result in no additional horizontal force, $F(t) \equiv 0$, and hence the leading-order velocity profile \eqref{Eq.OuterVelocity} is a uniform translation. The translation velocity is given by the far-field behaviour \eqref{Eq.LeadingResult} of the inner solution, so at leading order $\uout \approx -\omega_0\beta_0 = -0.3168$, which is independent of time.

The first corrections in the outer solution then come in at $\O(t) = \O(h_0)$. Time integration of the mass conservation equation \eqref{Eq.MassConservationScaled} yields the result
\begin{equation}\label{Eq.FreeOuterSol}
  h(x,t) = h_s(x) - t \uout h_s'(x) + \O(h_0^2) = h_s(x - \uout t) + \O(h_0^2),
\end{equation} 
which reveals that the first-order correction simply represents a translation of the initial profile $h_s(x)$ by the steady leading-order velocity $\uout < 0$. (In fact, it can be shown that to all orders in $h_0$, the velocity profile is spatially uniform and hence the drop is undergoing pure translation, with deformation only occuring in the bridge region $x = \O(h_0)$.)

In order to match with the bridge region, we substitute $x = h_0 \calX$ into \eqref{Eq.FreeOuterSol} and expand in powers of $h_0$, making use of $t \uout \approx -\omega_0 \beta_0  t \approx -\beta_0 h_0$. This yields
\begin{equation}
  h = h_0 \left[ \left(\calX - h_0\frac{\calX^2}{2}\right) + \beta_0 \left(1 - h_0 \calX\right) + \dots \right],
\end{equation}
Comparing this to the definition of the inner variables (\ref{Eq.InnerDef}), we now identify the appropriate far-field behaviour of the inner solution to be
\begin{equation}
  \calH_1(\calX) \sim -\frac{\calX^2}{2} - \beta_0 \calX + \dots \quad \text{as} \quad \calX \to \infty.
\end{equation}
This yields the condition $\calH_1'' \to -1$ as $\calX \to \infty$ that we anticipated in \S\ref{Sec.InnerFree}, which was imposed to obtain the solution \eqref{Eq.CorrectionResultFree}.

The resulting expression for the coalescence velocity is, from \eqref{Eq.LeadingResult} and \eqref{Eq.CorrectionResultFree},
\begin{equation}\label{Eq.VResultFree}
\dot h_0( t) = 0.5525 - 0.7625 h_0 + \O(h_0^2),
\end{equation}
predicting a linear correction to the coalescence velocity. This prediction is tested quantitatively in comparison to numerical simulations in figure \ref{fig:figure5}. The result (\ref{Eq.VResultFree}) is shown as a dashed line labeled ``free", while the solid lines are numerics that were initialised at three different initial bridge heights $h_{0i}$. After a short time, the numerical curves for free lenses rapidly converge to the predicted asymptotics (\ref{Eq.VResultFree}). Note that in typical experimental conditions where $h_0 \sim 10^{-3}\cdots 10^{-2}$, the coalescence velocity is very close to the leading order value $V_0$.

\begin{figure}
	\centering
	\includegraphics{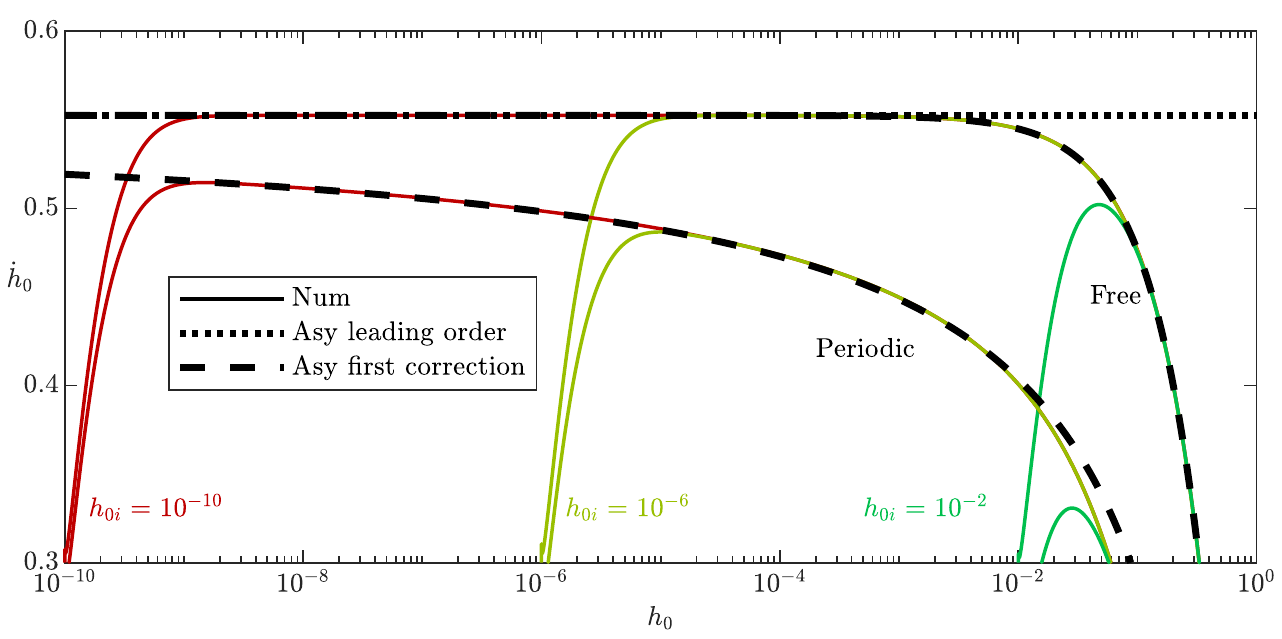} 
	\caption{The dependence of the coalescence velocity $V=\dot h_0$ on the bridge height $h_0$, for both free-floating \eqref{Eq.Free} and periodic \eqref{Eq.Periodic} lenses. Numerical results for three different values of the initial bridge height $h_{0i}$ are shown (solid lines). The asymptotic results for the free-floating and periodic lenses are given (dashed lines), respectively, by \eqref{Eq.VResultFree} and \eqref{Eq.VResultPeriodic}.}
	\label{fig:figure5}
\end{figure}

\subsection{The case of periodic lenses} 
\label{Sec.OuterPeriodic}

The periodic lenses come with the symmetry boundary condition \eqref{Eq.Periodic} on the velocity profile \eqref{Eq.OuterVelocity}. This yields $B(t)\equiv 0$ and hence
\begin{equation}\label{Eq.OuterVelocityPeriodic}
\uout = \dfrac{F(t)}{4} \ln{\dfrac{x}{2-x}}.
\end{equation}
This logarithmic velocity at small $x$ can indeed be matched to the velocity at large $\xi$ from the inner region calculated in \S\ref{Sec.InnerPeriodic}. Specifically, we equate \eqref{Eq.OuterVelocityPeriodic} (taking the limit $x \ll 1$) with the far-field inner velocity profile \eqref{Eq.LogFarfield} (using $u = V\calU$ and $\calX = x/h_0$),}
\begin{equation}\label{Eq.MatchVelocity}
\dfrac{F(t)}{4} \ln{\dfrac{x}{2}} = \dfrac{F(t)}{4} \ln{\dfrac{x}{h_0}} - \omega \beta,
\end{equation}
where it is noted that $\omega$ and $\beta$ are expanded in $F$ themselves. 

Equation (\ref{Eq.MatchVelocity}) finally enables us to express $F$ in terms of $h_0$, which to leading order gives the result
\begin{equation}\label{Eq.FSol}
  F = \frac{4\omega\beta}{\ln(2/h_0)} = \frac{4\omega_0\beta_0}{\ln(2/h_0)} + \O(\ln(2/h_0)^{-2}).
\end{equation}
This confirms the slow, logarithmic decay of the force term, suggesting strong corrections with respect to the leading order similarity solution. 
Most importantly, this leads us to the sought-after coalescence velocity for periodic drops:
\begin{equation}\label{Eq.VResultPeriodic}
 	\dot h_0 = \omega_0 + \omega_1 F  + \O(F^2) = \omega_0 + \frac{\hat\omega_1}{\ln(2/h_0)} + \O(\ln(2/h_0)^{-2}),
\end{equation}
where $\omega_0 = 0.5525$, $\omega_1 = -0.6227$ and $\hat\omega_1 = 4\omega_0\beta_0\omega_1 = 0.7892$.  
The $\O(\ln(2/h_0)^{-2})$ correction to the coalescence velocity is calculated in appendix \ref{Sec.Appendix}.

Once again, the asymptotic prediction is tested quantitatively in comparison to numerical simulations in figure \ref{fig:figure5}. The result (\ref{Eq.VResultPeriodic}) is shown as a dashed line labeled ``periodic", while the solid lines are numerics that were initialised at three different initial heights $h_{0i}$; again the numerics are in excellent agreement with the asymptotics. We notice a dramatic difference between the coalescence velocities for periodic drops as compared to the free drops, owing to the logarithmic decay of $F$. At values where $h_0 \sim 10^{-3}\cdots 10^{-2}$, the coalescence velocity for periodic drops is significantly below the leading order value $V_0$.

\section{Conclusion}\label{Sec.Conclusion}

In the present work, we have analysed the coalescence dynamics of viscous liquid lenses using both matched asymptotics and numerical simulations. We restricted our attention to the case where the flow in the bath is negligible, which is a consistent approximation for lenses of high viscosity, and where the contact angles are small to enable a slender thin-sheet description. In addition, following a previously used assumption in coalescence, we treated the problem as quasi-two-dimensional. The common scenario in coalescence and pinch-off, is that the flow remains localised into the narrow bridge region, while the far field remains stationary. Here we have found that this is not the case for viscous lens coalescence, and demonstrated that the bridge region affects the global dynamics, using both matched asymptotics and numerical simulations.

For freely floating two-dimensional viscous lenses, as soon as the lenses start coalescing there is a motion of the outer contact line towards the point of coalescence. In dimensional form, the growth rate of the bridge height $\bar{h}_0(\bar t) = \bar{h}(\bar x = 0, \bar{t})$ \eqref{Eq.VResultFree}, and the horizontal translation velocity of the drop's center of mass $\bar{u}_o$  (\ref{Eq.LeadingResult},\ref{Eq.CorrectionResultFree}) can be written as
\begin{subequations}\label{Eq.VResultFreeDimensional}
\begin{align}
\bar V = \dot{\bar{h}}_0 &= \left(0.5525 + 0.7625 \frac{\bar{h}_s''(0)\bar{h}_0}{\theta^2} + \dots\right) \frac{\gamma\theta^2}{\eta}, \\
\bar U = \bar{u}_o &= -\left(0.3168 - 0.5500 \frac{\bar{h}_s''(0)\bar{h}_0}{\theta^2} + \dots\right) \frac{\gamma\theta}{\eta}.
\end{align} 
\end{subequations}
Here, we have eliminated the lens radius $R$ in favour of the second derivative $\bar{h}_s''(0) = -\theta/R$ of the initial condition \eqref{Eq.Initial}, in order to highlight that the coalescence process depends only on the local shape of the lens near the bridge.

For a periodic array of two-dimensional viscous lenses, the centers of mass of the drops do not move due to symmetry. Instead, a horizontal force is generated that resists the inward motion towards the bridge, which results in a logarithmic velocity profile in the lens. In dimensional form, the growth rate of the bridge \eqref{Eq.VResultPeriodic} and the additional force generated \eqref{Eq.FSol} are then given by
\begin{align}\label{Eq.VResultPinnedDimensional}
  \bar V = \dot{\bar{h}}_0 = \left(0.5525 + \frac{0.7892}{\ln(2R\theta/\bar{h}_0)} + \dots\right) \frac{\gamma\theta^2}{\eta},  \quad
	\bar{F} = \left(\frac{1.267}{\ln(2R\theta/\bar{h}_0)} + \dots\right)\gamma \theta^2.
\end{align}
A further, second-order, correction is calculated in appendix \ref{Sec.Appendix}. It is important to note that the corrections are logarithmic in time, so these can be significant even at the very early stages of coalescence. In fact, logarithmic corrections also arise for viscous coalescence of freely suspended drops \citep{hopper_1990, Eggers1999}. The structure of the problem is, however, different. In the freely suspended case the dynamics is $ h\sim t \ln t$, which in contrast to (\ref{Eq.VResultPinnedDimensional}) does not exhibit a finite velocity at early times.

It is of interest to compare these two-dimensional, slender predictions to the experiments shown in figure \ref{fig:exp}(c). In \citet{Hack2020} it was already shown that the experimental vertical coalescence velocity $\bar V$ was in very good quantitative agreement with the leading-order prediction of (\ref{Eq.VResultFreeDimensional}). Indeed, for freely floating drops, the higher-order terms are expected to be negligible in the experimental range. The current theory predicts a ratio of horizontal to vertical velocity $\bar U/\bar V \approx 0.57/\theta$ for the freely moving drops, which for the experimental contact angles amounts to $\bar U/\bar V  \approx 1.1$. The experimentally measured ratio in figure \ref{fig:exp}(c) was found $\bar U/\bar V \approx 1.7$, which implies an even stronger center of mass motion than predicted. This quantitative disparity could be due to three-dimensional effects, or due to the fact that the contact angles are not very small. Still, our theory offers an explanation for the appearance of a center of mass motion for viscous lenses, and provides the relevant scaling laws. This center of mass motion is not observed for inertial drops, which is in accordance with the decaying velocity in the inertial similarity solution \citep{Hack2020}. For future experiments, it might be of interest to study coalescence while the centers of the drop are prevented from translating towards each other (e.g. by attaching the drops to fixed capillaries). In that case, we expect the appearance of significant (logarithmic) corrections to the coalescence velocity.

Besides three-dimensional effects, it is also of interest to discuss the influence of gravity. Gravity becomes important if the (dimensional) radius $R$ of the lenses is no longer small compared with the capillary length $\ell_c = \sqrt{\gamma/\Delta \rho\,g}$, where $\Delta \rho$ is the difference in density between the lenses and either of the external fluids. In this case, we expect gravity to flatten the static lens profile $h_s(x)$ \eqref{Eq.Initial}, but near the rim of the lens, when $x \ll \ell_c$, gravity becomes negligible and the capillary thin-sheet equations \eqref{Eq.GoverningScaled} and contact angle $\theta$ are recovered, and hence the leading-order inner solution \eqref{Eq.LeadingResult} will still hold. For (two-dimensional) free lenses, which can undergo uniform translation even in the presence of gravity, the $\O(h_0)$ correction \eqref{Eq.VResultFreeDimensional} also holds, but with a modified value of $h_s''(0)$ that simply follows from the static drop. For periodic lenses, the correction becomes more involved as the outer velocity profile \eqref{Eq.OuterVelocity} is affected by the change in $h_s$.

More generally, the dynamical structure of the problem bears a strong similarity with drop spreading on a rigid substrate, where the motion of the contact line also induces a weak flow on the scale of the drop \citep{bonn2009wetting}. The spreading velocity exhibits a logarithmic dependence on the scale separation between drop size and the characteristic scale of the contact line in that case too. An important difference, however, is that for drop spreading the universal leading-order similarity solution for the inner problem captures the phenomenon, as it has only algebraically small corrections. Here we found that for viscous lens coalescence, the corrections are themselves only logarithmically small, and therefore can be significant.

\medskip
We are grateful to J. Eggers for discussions. We would also like to thank the referees for all their very valuable comments and suggestions, and in particular for highlighting the different large-scale boundary conditions. W.T. acknowledges support by the European Union’s Horizon 2020 research and innovation programme under the Marie Sklodowska-Curie grant agreement No 722497 - LubISS. M. A. H. acknowledges support from an Industrial Partnership Programme of the Netherlands Organisation for Scientific Research (NWO), cofinanced by Canon Production Printing Netherlands B.V., University of Twente, and Eindhoven University of Technology. C.D. and J.H.S. acknowledge support from NWO through VICI Grant No. 680-47-632.\\

\smallskip
\textbf{Declaration of Interests:} The authors report no conflict of interest.

\appendix

\section{Details of the inner solution for non-zero $F$}\label{Sec.Appendix}

We calculate both the $\O(F)$ and $\O(F^2)$ corrections to the leading-order result \eqref{Eq.LeadingResult} by expanding 
\begin{gather}
  \calH = \calH_0 + F \calH_1 + F^2 \calH_2 + \frac{h_0}{\dot h_0}\dot F \calH_T + \O(F^3),
\end{gather}
and similarly for $\calU$, $\omega$ and $\beta$. Note that, since we anticipate that $F = \O(1/\ln t)$, the terms with subscript $2$ and $T$ are both $\O(F^2)$.

We substitute the expansion into the equations \eqref{inner_similarity}--\eqref{Eq.InnerBC}, together with the matching condition $\mathcal{H}'(\infty) = 1$ which is accurate to all orders in $F$, and identify coefficients.
The resulting governing equations for the corrections $\calH_\Delta$, $\calU_\Delta$ and $\omega_\Delta$, where $\Delta = 1, 2, T$, are given by
\begin{subequations}
\begin{gather}
  \begin{pmatrix}
	\calH_\Delta - \calX \calH_\Delta' + (\calH_0 \calU_\Delta + \calH_\Delta \calU_0)' \\
	\calH_0 \calH_\Delta'' + \calH_\Delta \calH_0'' - \calH_0' \calH_\Delta' + 4 (\omega_0 \calH_0 \calU_\Delta' + \omega_0 \calH_\Delta \calU_0' + \omega_\Delta \calH_0 \calU_0')
	\end{pmatrix} = {}\\
	{} = 
	\underbrace{\begin{pmatrix}
	0 \\ 1
	\end{pmatrix}}_{\Delta = 1}, 
	\underbrace{\begin{pmatrix}
	-(\calH_1 \calU_1)' \\ -\calH_1 \calH_1'' + \tfrac12 \calH_1'^2 - 4(\omega_0 \calH_1 \calU_1' + \omega_1 \calH_0 \calU_1' + \omega_1 \calH_1 \calU_0')
	\end{pmatrix}}_{\Delta = 2},
	\underbrace{\begin{pmatrix}
	  - \calH_1 \\ 0
	\end{pmatrix}}_{\Delta = T},
\end{gather}
\end{subequations}
and the boundary conditions are $\calH_\Delta'(0) = \calU_\Delta(0) = \calH_\Delta(0) = \calH_\Delta'(\infty) = 0$. 

Solving these equations numerically using Mathematica yields the results
\begin{align}
  \omega_0 &= 0.5525, & \omega_1 &= -0.6227, & \omega_2 &= -0.1267, & \omega_T &= 0.3028, \\ \beta_0 &= 0.5734, & \beta_1 &= 0.7079,& \beta_2 &= 0.8728, & \beta_T &= -0.5277,
\end{align}
for the coefficients in the far-field behaviour \eqref{Eq.LogFarfield}.

We can then use the matching \eqref{Eq.FSol} to obtain the results
\begin{subequations}
\begin{align}
 F &= \frac{4(\omega_0 + \omega_1 F + \O(F^2))(\beta_0 + \beta_1 F + \O(F^2))}{\ln(2/h_0)}\\
   &= \frac{4\omega_0\beta_0}{\ln(2/h_0)} + \frac{16 \omega_0\beta_0(\omega_0 \beta_1 + \omega_1 \beta_0)}{\ln(2/h_0)^2} + \O(\ln(2/h_0)^{-3}),
\end{align}
\end{subequations}
and
\begin{subequations}
\begin{align}
	\dot h_0 &= \omega_0 + \omega_1 F + \omega_2 F^2 + \omega_T (h_0/\dot h_0) \dot F + \O(F^3) = \\
	&= \omega_0 +	\frac{\hat\omega_1}{\ln(2/h_0)} + \frac{\hat\omega_2}{\ln(2/h_0)^2} + \O(\ln(2/h_0)^{-3}) = \\
	&= \omega_0 + \frac{\hat\omega_1}{\ln(c/h_0)} + \O(\ln(2/h_0)^{-3}),
\end{align}
\end{subequations}
where $\omega_0 = 0.5525$, $\hat\omega_1 = 4\omega_0\beta_0 \omega_1 = 0.7892$, and 
\begin{subequations}
\begin{align}
\hat\omega_2 &= 16\omega_0^2\beta_0^2 \omega_2 + 16\omega_0\beta_0(\omega_0\beta_1 + \omega_1\beta_0)\omega_1 + 4\omega_0\beta_0 \omega_T = 0.07272, \\
c &= 2\exp(-\hat\omega_2/\hat\omega_1) = 2.19.
\end{align}
\end{subequations}
The second-order coefficient $\hat\omega_2$ is coincidentally quite small when $\dot h_0$ is expanded in terms of $\ln(2/h_0)$, so including the second-order correction in figure \ref{fig:figure5} would not have a noticeable effect. However, when $\dot h_0$ is expanded in terms of e.g.\ $\ln(1/h_0)$ or $\ln(1/t)$, the second-order correction yields a significant improvement.

\bibliographystyle{jfm}

\bibliography{asymptoticsJFM}

\begin{thebibliography}{25}
\expandafter\ifx\csname natexlab\endcsname\relax\def\natexlab#1{#1}\fi
\def\au#1{#1} \def\ed#1{#1} \def\yr#1{#1}\def\at#1{#1}\def\jt#1{\textit{#1}}
  \def\bt#1{#1}\def\bvol#1{\textbf{#1}} \def\vol#1{#1} \def\pg#1{#1}
  \def\publ#1{#1}\def\arxiv#1{#1}\def\org#1{#1}\def\st#1{\textit{#1}}

\bibitem[Aarts {\em et~al.\/}(2005)Aarts, Lekkerkerker, Guo, Wegdam \&
  Bonn]{Aarts2005}
{\sc \au{Aarts, D. G. A.~L.}, \au{Lekkerkerker, H. N.~W.}, \au{Guo, H.},
  \au{Wegdam, G.~H.} \& \au{Bonn, D.}} \yr{2005}  \at{Hydrodynamics of droplet
  coalescence}.  \jt{Phys. Rev. Lett.}  \bvol{95},  \pg{164503}.

\bibitem[Anand {\em et~al.\/}(2012)Anand, Paxson, Dhiman, Smith \&
  Varanasi]{Anand2012}
{\sc \au{Anand, S.}, \au{Paxson, A.~T.}, \au{Dhiman, R.}, \au{Smith, J.~D.} \&
  \au{Varanasi, K.~K.}} \yr{2012}  \at{Enhanced condensation on
  lubricant-impregnated nanotextured surfaces}.  \jt{ACS Nano}  \bvol{6},
  \pg{10122--10129}.

\bibitem[Billingham \& King(2005)]{Billingham2005}
{\sc \au{Billingham, J.} \& \au{King, A.~C.}} \yr{2005}
  \at{Surface-tension-driven flow outside a slender wedge with an application
  to the inviscid coalescence of drops}.  \jt{J. Fluid Mech.}  \bvol{533},
  \pg{193--221}.

\bibitem[Bonn {\em et~al.\/}(2009)Bonn, Eggers, Indekeu, Meunier \&
  Rolley]{bonn2009wetting}
{\sc \au{Bonn, D.}, \au{Eggers, J.}, \au{Indekeu, J.}, \au{Meunier, J.} \&
  \au{Rolley, E.}} \yr{2009}  \at{Wetting and spreading}.  \jt{Rev. Mod. Phys.}
   \bvol{81}~(2),  \pg{739}.

\bibitem[Burton \& Taborek(2007)]{Burton2007}
{\sc \au{Burton, J.~C.} \& \au{Taborek, P.}} \yr{2007}  \at{Role of
  dimensionality and axisymmetry in fluid pinch-off and coalescence}.
  \jt{Phys. Rev. Lett.}  \bvol{98},  \pg{224502}.

\bibitem[Delabre \& Cazabat(2010)]{Delabre2010}
{\sc \au{Delabre, U.} \& \au{Cazabat, A.-M.}} \yr{2010}  \at{Coalescence driven
  by line tension in thin nematic films}.  \jt{Phys. Rev. Lett.}  \bvol{104},
  \pg{227801}.

\bibitem[Duchemin {\em et~al.\/}(2003)Duchemin, Eggers \&
  Josserand]{Duchemin2003}
{\sc \au{Duchemin, L.}, \au{Eggers, J.} \& \au{Josserand, C.}} \yr{2003}
  \at{Inviscid coalescence of drops}.  \jt{J. Fluid Mech.}  \bvol{487},
  \pg{167--178}.

\bibitem[Eddi {\em et~al.\/}(2013)Eddi, Winkels \& Snoeijer]{Eddi2013}
{\sc \au{Eddi, A.}, \au{Winkels, K.~G.} \& \au{Snoeijer, J.~H.}} \yr{2013}
  \at{Influence of droplet geometry on the coalescence of low viscosity drops}.
   \jt{Phys.~Rev.~Lett.}  \bvol{111},  \pg{144502}.

\bibitem[Eggers \& Fontelos(2015)]{Eggers2015}
{\sc \au{Eggers, J.} \& \au{Fontelos, M.~A.}} \yr{2015} {\em Singularities:
  Formation, Structure, and Propagation\/}.  \publ{Cambridge University Press}.

\bibitem[Eggers {\em et~al.\/}(1999)Eggers, Lister \& Stone]{Eggers1999}
{\sc \au{Eggers, J.}, \au{Lister, J.~R.} \& \au{Stone, H.~A.}} \yr{1999}
  \at{Coalescence of liquid drops}.  \jt{J. Fluid Mech.}  \bvol{401},
  \pg{293--310}.

\bibitem[Erneux \& Davis(1993)]{Erneux1993}
{\sc \au{Erneux, T.} \& \au{Davis, S.~H.}} \yr{1993}  \at{Nonlinear rupture of
  free films}.  \jt{Phys. Fluids A}  \bvol{5},  \pg{1117--1122}.

\bibitem[de~Gennes {\em et~al.\/}(2004)de~Gennes, Brochard-Wyart \&
  Qu{\'e}r{\'e}]{deGennes2004}
{\sc \au{de~Gennes, P.-G.}, \au{Brochard-Wyart, F.} \& \au{Qu{\'e}r{\'e}, D.}}
  \yr{2004} {\em Capillarity and Wetting Phenomena: Drops, Bubbles, Pearls,
  Waves\/}.  \publ{Springer}.

\bibitem[Hack {\em et~al.\/}(2018)Hack, Costalonga, Segers, Karpitschka,
  Wijshoff \& Snoeijer]{Hack2018}
{\sc \au{Hack, M.~A.}, \au{Costalonga, M.}, \au{Segers, T.}, \au{Karpitschka,
  S.}, \au{Wijshoff, H.} \& \au{Snoeijer, J.~H.}} \yr{2018}  \at{Printing
  wet-on-wet: Attraction and repulsion of drops on a viscous film}.  \jt{Appl.
  Phys. Lett.}  \bvol{113},  \pg{183701}.

\bibitem[Hack {\em et~al.\/}(2020)Hack, Tewes, Xie, Datt, Harth, Harting \&
  Snoeijer]{Hack2020}
{\sc \au{Hack, M.~A.}, \au{Tewes, W.}, \au{Xie, Q.}, \au{Datt, C.}, \au{Harth,
  K.}, \au{Harting, J.} \& \au{Snoeijer, J.~H.}} \yr{2020}  \at{Self-similar
  liquid lens coalescence}.  \jt{Phys.~Rev.~Lett.}  \bvol{124}~(19),
  \pg{194502}.

\bibitem[Hern\'andez-S\'anchez {\em et~al.\/}(2012)Hern\'andez-S\'anchez,
  Lubbers, Eddi \& Snoeijer]{HernandezSanchez2012}
{\sc \au{Hern\'andez-S\'anchez, J.~F.}, \au{Lubbers, L.~A.}, \au{Eddi, A.} \&
  \au{Snoeijer, J.~H.}} \yr{2012}  \at{Symmetric and asymmetric coalescence of
  drops on a substrate}.  \jt{Phys.~Rev.~Lett.}  \bvol{109},  \pg{184502}.

\bibitem[Hopper(1990)]{hopper_1990}
{\sc \au{Hopper, R.~W.}} \yr{1990}  \at{Plane stokes flow driven by capillarity
  on a free surface}.  \jt{J. Fluid Mech.}  \bvol{213},  \pg{349–375}.

\bibitem[Kamp {\em et~al.\/}(2016)Kamp, Villwock \& Kraume]{Kamp2016}
{\sc \au{Kamp, J.}, \au{Villwock, J.} \& \au{Kraume, M.}} \yr{2016}  \at{Drop
  coalescence in technical liquid/liquid applications: a review on experimental
  techniques and modeling approaches}.  \jt{Rev. Chem. Eng.}  \bvol{33},
  \pg{1--47}.

\bibitem[Lee {\em et~al.\/}(2012)Lee, Kang, Yoon \& Yarin]{Lee2012}
{\sc \au{Lee, M.~W.}, \au{Kang, D.~K.}, \au{Yoon, S.~S.} \& \au{Yarin, A.~L.}}
  \yr{2012}  \at{Coalescence of two drops on partially wettable substrates}.
  \jt{Langmuir}  \bvol{28},  \pg{3791--3798}.

\bibitem[Narge {\em et~al.\/}(2008)Narge, Beysens \& Pomeau]{Narhe2008}
{\sc \au{Narge, R.~D.}, \au{Beysens, D.~A.} \& \au{Pomeau, Y.}} \yr{2008}
  \at{Dynamics drying in the early-stage coalescence of droplets sitting on a
  plate}.  \jt{Eur. Phys. Lett.}  \bvol{81},  \pg{46002}.

\bibitem[Paulsen {\em et~al.\/}(2011)Paulsen, Burton \& Nagel]{Paulsen2011}
{\sc \au{Paulsen, J.~D.}, \au{Burton, J.~C.} \& \au{Nagel, S.~R.}} \yr{2011}
  \at{Viscous to inertial crossover in liquid drop coalescence}.  \jt{Phys.
  Rev. Lett.}  \bvol{106},  \pg{114501}.

\bibitem[Ristenpart {\em et~al.\/}(2006)Ristenpart, McCalla, Roy \&
  Stone]{Ristenpart2006}
{\sc \au{Ristenpart, W.~D.}, \au{McCalla, P.~M.}, \au{Roy, R.~V.} \& \au{Stone,
  H.~A.}} \yr{2006}  \at{Coalescence of spreading droplets on a wettable
  substrate}.  \jt{Phys. Rev. Lett.}  \bvol{97},  \pg{064501}.

\bibitem[Shaw(2003)]{Shaw2003}
{\sc \au{Shaw, J.~M.}} \yr{2003}  \at{A microscopic view of oil slick break-up
  and emulsion formation in breaking waves}.  \jt{Spill Sci. Technol. Bull.}
  \bvol{8},  \pg{491--501}.

\bibitem[Smith {\em et~al.\/}(2013)Smith, Dhiman, Anand, Reza-Garduno, Cohen,
  McKinley \& Varanasi]{Smith2013}
{\sc \au{Smith, J.~D.}, \au{Dhiman, R.}, \au{Anand, S.}, \au{Reza-Garduno, E.},
  \au{Cohen, R.~E.}, \au{McKinley, G.~H.} \& \au{Varanasi, K.~K.}} \yr{2013}
  \at{Droplet mobility on lubricant-impregnated surfaces}.  \jt{Soft Matter}
  \bvol{9},  \pg{1772--1780}.

\bibitem[Thoroddsen {\em et~al.\/}(2007)Thoroddsen, Qian, Etoh \&
  Takehara]{Thoroddsen2007}
{\sc \au{Thoroddsen, S.~T.}, \au{Qian, B.}, \au{Etoh, T.~G.} \& \au{Takehara,
  K.}} \yr{2007}  \at{The initial coalescence of miscible drops}.  \jt{Phys.
  Fluids}  \bvol{19},  \pg{072110}.

\bibitem[Ting \& Keller(1990)]{Ting1990}
{\sc \au{Ting, L.} \& \au{Keller, J.~B.}} \yr{1990}  \at{Slender jets and thin
  sheets with surface tension}.  \jt{SIAM J. Appl. Math.}  \bvol{50},
  \pg{1533--1546}.

\end{thebibliography}

\end{document}